\documentclass[12pt]{article}
\usepackage{a4wide}
\usepackage{epsfig}
\usepackage{amsmath}

\usepackage{latexsym}
\usepackage{amssymb}
\usepackage{multicol}

\newcommand{\half}{{\textstyle\frac{1}{2}}}

\newcommand{\fourth}{{\textstyle\frac{1}{4}}}
\newcommand{\threehalf}{{\textstyle\frac{3}{2}}}

\newcommand{\dd}{\mbox{{\rm d}}}
\newcommand{\GF}{G_{\rm F}}

\catcode`@=11
\def\citer{\@ifnextchar [{\@tempswatrue\@citexr}{\@tempswafalse\@citexr[]}}

\def\@citexr[#1]#2{\if@filesw\immediate
  \write\@auxout{\string\citation{#2}}\fi
  \def\@citea{}\@cite{\@for\@citeb:=#2\do
    {\@citea\def\@citea{--\penalty\@m}\@ifundefined
       {b@\@citeb}{{\bf ?}\@warning
       {Citation `\@citeb' on page \thepage \space undefined}}%
\hbox{\csname b@\@citeb\endcsname}}}{#1}}
\catcode`@=12

\begin{document}

\begin{center}
{\Large \bf Three-Neutrino MSW Effect \\[6mm]
            and the LNW Mass Matrix}
\end{center}
\vspace{8mm}
\begin{center}
Per Osland$^{a}$
 \ \ and \ \
Tai Tsun Wu$^{b}$
\\
\vspace{1cm}
$^{a}${\em
Department of Physics, University of Bergen, \\
      All\'{e}gaten 55, N-5007 Bergen, Norway}
\\
\vspace{.3cm}
$^{b}${\em
Gordon McKay Laboratory, Harvard University, \\
Cambridge, Massachusetts 02138, U.S.A.}
\\
{\em and} \\
{\em
Theoretical Physics Division, CERN,
CH-1211 Geneva 23, Switzerland}
\\
\end{center}
\hspace{10mm}
\begin{abstract}
We review recent work on analytical solutions to the MSW equations
for three neutrino flavours, for exponential and linear potentials.
An application to a particular mass matrix is also discussed.
The three neutrino masses are determined, respectively, to be
0.001--0.004, and roughly 0.01 and 0.05~eV.
\end{abstract}
\vspace{.3cm}

\begin{multicols}{2}
\section{Introduction}
In this paper we shall review some recent results on analytical solutions
of the Mikheyev--Smirnov--Wolfenstein (MSW) effect \cite{MSW}
for the propagation of three neutrino flavours.
For two model densities which are relevant to the neutrino propagation
in the sun, such results have been obtained.
These are the exponential density \cite{Osland:2000et} and
the linear density \cite{Lehmann:2000ey}.
(The case of a constant density, which is of some relevance
for propagation through the Earth, has also been studied recently
\cite{Ohlsson:2000xb}.)

For the exponential density, the solutions for the three neutrino
wave functions \cite{Osland:2000et} can be expressed in terms of
generalized hypergeometric functions, ${}_2F_2$ and ${}_3F_1$.
For the linear density, the solutions can be expressed as a Fourier
transform of a rather simple expression, which, in the case $N=2$
(two flavours) reduces to the well-known parabolic cylinder functions
or confluent hypergeometric functions \cite{two-nu}.

We also briefly discuss the application of the results to the
Lehmann--Newton--Wu (LNW) mass matrix \cite{Lehmann:1996br,Osland:2000bh}.

\section{Exponential density}
Neutrino propagation through a medium where the electron neutrino
(denoted $\phi_1(r)$) interacts differently from the others,
is governed by the equation
\begin{multline}
\label{Eq:Schr-1}
i\frac{\dd}{\dd r}
\begin{bmatrix}
\phi_1(r) \\ \phi_2(r) \\ \phi_3(r)
\end{bmatrix}
=\left(
\begin{bmatrix}
D(r) & 0 & 0 \\
0 & 0 & 0 \\
0 & 0 & 0 \\
\end{bmatrix} \right. \\
\left.
+\frac{1}{2p}
\begin{bmatrix}
M^2_{11} & M^2_{12} & M^2_{13} \\
M^2_{21} & M^2_{22} & M^2_{23} \\
M^2_{31} & M^2_{32} & M^2_{33}
\end{bmatrix}
\right)
\begin{bmatrix}
\phi_1(r) \\ \phi_2(r) \\ \phi_3(r)
\end{bmatrix} ,
\end{multline}
where the mass matrix is real and symmetric,
$M^2_{ji}=M^2_{ij}\equiv (M^2)_{ij}$. Here,
$D(r)=\sqrt{2}\GF N_e(r)$, with $\GF$ the Fermi constant and
$N_e(r)$ the solar electron density.

For the sun, the density \cite{BBP-98}
is well approximated by an exponential,
\begin{equation}
N_e(r) = N_e(0)\, e^{-r/r_0},\quad
r_0\simeq 0.1\times R_\odot .
\end{equation}

It is convenient to introduce a new radial variable:
$u=r/r_0+u_0$, and perform a rotation on the second and third components,
\begin{multline}
\begin{bmatrix}
\cos\theta_0 & -\sin\theta_0 \\
\sin\theta_0 & \cos\theta_0
\end{bmatrix}
\frac{r_0}{2p}
\begin{bmatrix}
M^2_{22} & M^2_{23} \\
M^2_{32} & M^2_{33}
\end{bmatrix}  \\
\times
\begin{bmatrix}
\cos\theta_0 & \sin\theta_0 \\
-\sin\theta_0 & \cos\theta_0
\end{bmatrix}
=
\begin{bmatrix}
\omega_2 & 0 \\
0 & \omega_3
\end{bmatrix} .
\end{multline}
Eq.~(\ref{Eq:Schr-1}) then takes the form
\begin{equation}
\label{Eq:Schr-4}
i\frac{\dd}{\dd u}
\begin{bmatrix}
\psi_1(u) \\ \psi_2(u) \\ \psi_3(u)
\end{bmatrix}
=
\begin{bmatrix}
\omega_1+e^{-u} & \chi_2 & \chi_3 \\
\chi_2 & \omega_2 & 0 \\
\chi_3 & 0 & \omega_3
\end{bmatrix} \!
\begin{bmatrix}
\psi_1(u) \\ \psi_2(u) \\ \psi_3(u)
\end{bmatrix}.
\end{equation}

Let $\mu_1$, $\mu_2$ and $\mu_3$ be the eigenvalues of the $3\times3$
matrix
\begin{equation}
\begin{bmatrix}
\omega_1 & \chi_2 & \chi_3 \\
\chi_2 & \omega_2 & 0 \\
\chi_3 & 0 & \omega_3
\end{bmatrix} .
\end{equation}
These $\mu_j$ are the squares of the neutrino masses multiplied by
$r_0/(2p)$.
Together with $\omega_1$ and $\omega_2$ they control the evolution
of the $\psi_i$.

We again introduce a new variable,
$z=i e^{-u}$.
Then, the solutions to Eq.~(\ref{Eq:Schr-4}) can be expressed in terms of
solutions to
\begin{multline}
\label{Eq:Schr-6}
\biggl[
\left(z\frac{\dd}{\dd z}-i\mu_1\right)
\left(z\frac{\dd}{\dd z}-i\mu_2\right)
\left(z\frac{\dd}{\dd z}-i\mu_3\right)  \\
-z\left(z\frac{\dd}{\dd z}-i\omega_2\right)
  \left(z\frac{\dd}{\dd z}-i\omega_3\right)
\biggr]\psi=0,
\end{multline}
namely generalized hypergeometric functions \cite{Bateman-1}:
\begin{align}
&\psi^{(1)}
=e^{-i\mu_1u} \nonumber \\
&\times {}_2F_2\biggl[
\begin{matrix}
-i(\omega_2-\mu_1), & -i(\omega_3-\mu_1)\\
1-i(\mu_2-\mu_1),    & 1-i(\mu_3-\mu_1)
\end{matrix}
\bigg|ie^{-u} \biggr] \nonumber \\
&\psi^{(2)}
=e^{-i\mu_2u} \nonumber \\
&\times {}_2F_2\biggl[
\begin{matrix}
-i(\omega_2-\mu_2), & -i(\omega_3-\mu_2)\\
1-i(\mu_1-\mu_2),    & 1-i(\mu_3-\mu_2)
\end{matrix}
\bigg|ie^{-u} \biggr] \nonumber \\
&\psi^{(3)}
=e^{-i\mu_3u} \nonumber \\
&\times {}_2F_2\biggl[
\begin{matrix}
-i(\omega_2-\mu_3), & -i(\omega_3-\mu_3)\\
1-i(\mu_1-\mu_3),    & 1-i(\mu_2-\mu_3)
\end{matrix}
\bigg|ie^{-u} \biggr]
\end{align}
The ${}_2F_2$ can be defined in terms of the series expansions
\begin{equation}
\label{Eq:F22-series}
{}_2F_2\biggl[
\begin{matrix}
\alpha_1, & \alpha_2\\
\rho_1,   & \rho_2
\end{matrix}
\bigg|z \biggr]
=\sum_{k=0}^\infty
\frac{(\alpha_1)_k(\alpha_2)_k}{(\rho_1)_k(\rho_2)_k}\,
\frac{z^k}{k!}
\end{equation}
where $(\alpha)_k$ is a Pochhammer symbol,
$(\alpha)_k=\alpha(\alpha+1)\ldots(\alpha+k-1)$.
The solutions to Eq.~(\ref{Eq:Schr-4}) are thus
\begin{equation}
\label{Eq:psi_i}
\psi_i=C_1\psi_i^{(1)}+C_2\psi_i^{(2)}+C_3\psi_i^{(3)},
\end{equation}
where the constants $C_j$ are determined by the boundary
conditions: $\psi_1(r=0)=1$, $\psi_2(r=0)=0$, $\psi_3(r=0)=0$.

For neutrino masses and energies of physical interest, the
parameters and arguments of the ${}_2F_2$ become too large
for the series expansion to be useful.
A practical procedure is to adopt a stationary-phase approximation
for the ${}_2F_2$ of $\psi_i^{(3)}$,
and express the others by ${}_3F_1$ functions.
This procedure is outlined below.

Consider the ordinary differential equation for $_2F_2$ in the form
\begin{multline}
\label{Eq:Schr-7}
\biggl[
\left(z\,\frac{d}{dz}+\beta_1\right)
\left(z\,\frac{d}{dz}+\beta_2\right)
\left(z\,\frac{d}{dz}+\beta_3\right) \\
-z\left(z\,\frac{d}{dz}+\alpha_1\right)
  \left(z\,\frac{d}{dz}+\alpha_2\right)
\biggr]f=0.
\end{multline}
This can be converted to an ordinary differential equation
for ${}_3F_1$ by substituting $\hat z=z^{-1}$:
\begin{multline}
\label{Eq:Schr-8}
\biggl[
\left(\hat z\,\frac{d}{d\hat z}-\alpha_1\right)
\left(\hat z\,\frac{d}{d\hat z}-\alpha_2\right) \\
+\hat z\left(\hat z\,\frac{d}{d\hat z}-\beta_1\right)
       \left(\hat z\,\frac{d}{d\hat z}-\beta_2\right) \\
       \left(\hat z\,\frac{d}{d\hat z}-\beta_3\right)
\biggr]f=0 .
\end{multline}
Two solutions are of the form
\begin{equation}
\label{Eq:hat-f}
z^{\rm power}
{}_3F_1\left[\left.
\begin{matrix}
a_1, & a_2, & a_3\cr
b & &
\end{matrix}
\right|-z^{-1} \right] .
\end{equation}

The full solutions (\ref{Eq:psi_i}) can then be constructed
schematically (leaving out powers) as follows:
\begin{equation}
\psi_i=A_{i}\,{}_2F_2(1)+B_{i}\,{}_3F_1(1)
+C_{i}\,{}_3F_1(2) .
\end{equation}
The series expansion for ${}_3F_1$ has zero radius of convergence.
However, it can be expressed in terms of an integral
involving the familiar hypergeometric function ${}_2F_1$,
\begin{multline}
{}_3F_1\biggl[
\begin{matrix}
a_1, & a_2, & a_3 \cr
b &  &
\end{matrix}
\bigg|-x^{-1} \biggr] \\
=\frac{1}{\Gamma(a_1)}\,x^{a_1}
\int_0^\infty dt\, e^{-xt}\, t^{a_1-1}\nonumber \\
\times {}_2F_1(a_2,a_3;b;-t) \nonumber
\end{multline}

For large parameters and argument, the Pochhammer contour ${\cal P}$
is useful \cite{Osland:2000et}:
\begin{multline}
\label{Eq:F21-Pochhammer}
{}_2F_1(a,b;c;-t) \nonumber \\
=\frac{-\Gamma(c)e^{-i\pi c}}
        {4\Gamma(b)\Gamma(c-b)\sin\pi b\sin\pi(c-b)} \nonumber \\[5pt]
\mbox{}\times\int_{\cal P} s^{b-1}(1-s)^{c-b-1}(1+ts)^{-a}ds
\end{multline}
This approach leads to an accurate and efficient evaluation
of the three neutrino wave functions in terms of stationary
phase approximations to the ${}_2F_2$ and ${}_3F_1$ functions.
\section{Linear density}
The terminology ``linear electron density'' is used to mean that
$N_e(x)$ is a linear function of $x$.

\subsection{Two generations}

The case of two states and a linear potential has been studied
extensively, starting with Landau and Zener in the 1930s
and applied to neutrino mixing in the 1980s \cite{two-nu}.
After a suitable scaling and shift of the variable, one has
\begin{equation}
i\frac{d}{dt}
\begin{bmatrix}
\psi_1(t) \\ \psi_2(t)
\end{bmatrix}
=\begin{bmatrix}
 -t & a_2 \\
a_2 & 0
\end{bmatrix}
\begin{bmatrix}
\psi_1(t) \\ \psi_2(t)
\end{bmatrix} ,
\end{equation}
or more explicitly:
\begin{eqnarray}
\label{Eq:MSW-2-nu}
i\frac{d}{dt}\psi_1(t)
&=&-t\psi_1(t)+a_2\psi_2(t), \nonumber \\
\label{Eq:MSW-2-nu-last}
i\frac{d}{dt}\psi_2(t)
&=&a_2\psi_1(t) .
\end{eqnarray}
Elimination of $\psi_2$ gives
\begin{equation}
\label{Eq:MSW-2-nu-other}
\frac{d^2\psi_1(t)}{dt^2}-it\,\frac{d\psi_1(t)}{dt}
+(a_2^2-i)\psi_1(t)=0.
\end{equation}

The first-derivative term can be removed by taking
\begin{equation}
\psi_1(t)=e^{it^2/4}\,\phi_1(t) .
\end{equation}
Then the equation for $\phi_1(t)$ is
\begin{equation}
\label{Eq:MSW-2-nu-second-o}
\frac{d^2\phi_1(t)}{dt^2}+(\fourth t^2+a_2^2-\half i)\phi_1(t)=0 .
\end{equation}
Two linearly independent solutions of this
equation are the parabolic cylinder functions
\begin{equation}
D_\rho(\pm e^{i\pi/4}\,t)
\end{equation}
where $\rho=-ia_2^2-1$.

Parabolic cylinder functions are special cases of the confluent
hypergeometric function,
\begin{equation}
D_\rho(z)=2^{(\rho-1)/2}\, e^{-z^2/4}\, z\,
\Psi(\half-\half\rho,\threehalf;\half z^2)
\end{equation}
In terms of confluent hypergeometric functions $\Psi$ and $\Phi$:
\begin{eqnarray}
\label{Eq:psi-hyper}
\psi_1(t)
&=&t\bigl[C\,\Phi(1+\half ia_2^2,\threehalf;\half i t^2) \nonumber \\
&& +C'\,\Psi(1+\half ia_2^2,\threehalf;\half i t^2)\bigr]
\end{eqnarray}
Unfortunately, it is not clear how to generalize this approach to $N\ge3$.

Let us therefore consider an alternative solution to the $N=2$ case
\cite{Lehmann:2000ey}.
We start by writing
\begin{equation}
\label{Eq:capF-Fourier}
F(\zeta)=\frac{1}{2\pi}\int_{-\infty}^\infty dt\, e^{i\zeta t}\,
\psi_1(t) .
\end{equation}
Then it follows from Eq.~(\ref{Eq:MSW-2-nu-other}) that $F(\zeta)$
satisfies the first-order differential equation
\begin{equation}
-\zeta^2F(\zeta)-\frac{d}{d\zeta}[-i\zeta F(\zeta)]
+(a_2^2-i)F(\zeta)=0
\end{equation}
or
\begin{equation}
\label{Eq:capF-d.e.}
\frac{1}{F(\zeta)}\,\frac{dF(\zeta)}{d\zeta}
=\frac{i}{\zeta}(a_2^2-\zeta^2) .
\end{equation}
Integrating over $\zeta$, one finds
\begin{equation}
\label{Eq:capF}
F(\zeta)=\text{const.}\,e^{-i\zeta^2/2}\, \zeta^{ia_2^2}.
\end{equation}

With the notation [cf.\ Eq.~(\ref{Eq:A})]
\begin{equation}
\label{Eq:b-extras}
b_1=-\infty \quad \text{and} \quad b_{N+1}=+\infty ,
\end{equation}
the two solutions can be written as
\begin{equation}
\psi_1^{(n)}(t)=\int_{b_n}^{b_{n+1}} d\zeta\, e^{-i\zeta t}\,
e^{-i\zeta^2/2}\,|\zeta|^{ia_2^2}
\end{equation}
for $n=1,2$.
It can be shown that they are confluent hypergeometric functions
of the correct parameters and argument [i.e., identical to
Eq.~(\ref{Eq:psi-hyper})].

\subsection{General $N$}

This second approach outlined above has the advantage
that it can be generalized to an arbitrary number of neutrino
flavors \cite{Lehmann:2000ey}.
We start out by writing the equation analogous to (\ref{Eq:Schr-1})
in dimensionless standard form
\begin{equation}
\label{Eq:psi-Schr}
i\frac{d}{dt}\,\psi(t)=A(t)\psi(x) ,
\end{equation}
where
\begin{equation}
\label{Eq:psi}
\psi(t)=
\begin{bmatrix}
\psi_1(t) \\[2pt] \psi_2(t) \\[2pt] \psi_3(t) \\[2pt] \vdots \\[2pt]
\psi_N(t)
\end{bmatrix}
\end{equation}
and
\begin{equation}
\label{Eq:A}
A(t)=
\begin{bmatrix}
-t\,  & a_2\, & a_3\, & \hdots & a_N \\[1pt]
a_2   & b_2 & 0   & \hdots & 0 \\[1pt]
a_3   & 0   & b_3 & \hdots & 0 \\[1pt]
\vdots & \vdots   & \vdots &  & \vdots \\[1pt]
a_N & 0   & 0   & \hdots & b_N
\end{bmatrix} .
\end{equation}

There are two kinds of equations:
\begin{equation}
\label{Eq:psi1-5}
i\,\frac{d\psi_1(t)}{dt}=-t\,\psi_1(t)+\sum_{j=2}^N a_j\psi_j(t)
\end{equation}
and, for $k=2,3,4\ldots N$:
\begin{equation}
\label{Eq:psik-5}
\left(i\,\frac{d}{dt}-b_k\right)\psi_k(t)=a_k\psi_1(t) .
\end{equation}

One finds
\begin{equation}
\label{Eq:psi-n-sum}
\psi(t)=\sum_{n=1}^N C_n\,\psi^{(n)}(t) ,
\end{equation}
where \cite{Lehmann:2000ey}
\begin{multline}
\label{Eq:psi-n}
\psi^{(n)}(t)=\int_{b_n}^{b_{n+1}} d\zeta\, e^{-i\zeta t}\,
e^{-i\zeta^2/2}  \\
\times\left(\prod_{j=2}^N|\zeta-b_j|^{ia_j^2}\right)
\begin{bmatrix}
1 \\ \frac{a_2}{\zeta-b_2} \\ \frac{a_3}{\zeta-b_3} \\ \vdots \\
\frac{a_{N-1}}{\zeta-b_{N-1}} \\ \frac{a_N}{\zeta-b_N}
\end{bmatrix}
\end{multline}

While these solutions are reasonably simple, a numerical Fourier
transform is required.
As $t\to\pm\infty$, explicit expressions can be written out
for the different $\psi^{(n)}$.
However, these can not directly be used for imposing the boundary
conditions, since they would correspond to negative density at
$t\to -\infty$.

\section{The LNW mass matrix}

For quark mixing, it was found \cite{Lehmann:1996br}
that a particular, simple texture
for the $d$ ($d$, $s$, $b$) and $u$ ($u$, $c$, $t$) quark mass
matrices leads to an acceptable CKM matrix \cite{CKM}.
This same mass matrix has been applied to the case of three
neutrinos \cite{Osland:2000bh}, and rather good fits to the
atmospheric \cite{Super-K-prl} and solar
\citer{Super-K-sun,GALLEX} neutrino data have been obtained.

The mass matrix is assumed to have the form
\begin{equation}
M=\begin{pmatrix}
0  & d & 0 \\
d  & c & b \\
0  & b & a
\end{pmatrix}
\end{equation}
with $b^2=8c^2$.
The eigenvalues are given by $m_1$, $m_2$, and $m_3$, with $m_1\le m_3$.

In order to outline the diagonalization,
whereby $M=RM_{\rm diag}R^{\rm T}$,
let us introduce the notation
\begin{eqnarray}
S_1&\equiv& m_3-m_2+m_1, \nonumber \\
&=&a+c \nonumber \\
-S_2&\equiv&m_3m_2-m_3m_1+m_2m_1, \nonumber \\
&=&8c^2+d^2-ac \nonumber \\
-S_3&\equiv& m_1m_2m_3 \nonumber \\
&=&ad^2.
\end{eqnarray}
Then, a cubic equation for the parameter $a$ can be written as
\begin{equation}
9a^3-17S_1a^2+(8S_1^2+S_2)a-S_3=0.
\end{equation}
A physical solution requires $a$ real and positive.
This is equivalent to having three real solutions for $a$.
One of these is negative and two are positive.
At any point inside the allowed domain in the $m_1/m_3$--$m_2/m_3$
plane, there are thus two allowed solutions, denoted Solutions~1 and 2.

Let us consider first the atmospheric neutrino data.
The Super-Kamiokande results \cite{Super-K-prl} give
$\Delta m^2 \simeq (2-3)\times 10^{-3}$~eV, with
$\sin^2(2\theta)\simeq 1$.
The survival of muon neutrinos is given by
\begin{multline}
\label{Eq:atm-survival}
P_{\nu_\mu\rightarrow \nu_\mu}(t)
= 1
-4\biggl[U_{\mu1}^2U_{\mu2}^2\sin^2\left(\frac{\Delta m^2_{21}t}{4p}\right)
\\
+U^2_{\mu1}U^2_{\mu3}\sin^2\left(\frac{\Delta m^2_{31}t}{4p}\right)
\\
+U_{\mu2}^2U_{\mu3}^2\sin^2\left(\frac{\Delta m^2_{32}t}{4p}\right)
\biggr] ,
\end{multline}
where $U$ is the neutrino mixing matrix.
In the limit of $\Delta m^2_{21}t/4p \ll 1$ this simplifies,
and invoking further unitarity, one finds
\begin{equation}
P_{\nu_\mu\rightarrow \nu_\tau}(t)
\simeq 4U^2_{\mu3}U^2_{\tau3}\,
\sin^2\left(\frac{\Delta m^2_{32}t}{4p}\right) ,
\end{equation}
which suggests that one needs
$|U_{\mu3}U_{\tau3}|={\cal O}(1)$.
This can be achieved within the model (for both Solutions 1 and 2),
for $m_1\ll m_3$, with also $m_2$ small compared with $m_3$.
Furthermore, the data suggest that the scale $m_3$ must be
such that $m_3^2\simeq (2-3)\times 10^{-3}$~eV.

Fits to atmospheric data confirm this qualitative analysis.
Invoking also the solar Cl, Ga and Super-Kamiokande
neutrino data \citer{Super-K-sun,GALLEX},
one finds that both Solutions~1 and 2 give good fits for
$m_1\ll m_3$, with $m_2$ also small as compared with $m_3$.
Forming a $\chi^2$ from these different atmospheric and solar
survival probabilities, we found good fits \cite{Osland:2000bh},
with $m_3$ of the order of 0.05~eV, $m_2$ about 0.01~eV,
and $m_1\sim$ 0.001--0.004~eV.
In terms of the more conventional two-flavour analyses for
the solar-neutrino sector, these fits roughly correspond to
the large-mixing-angle solution.
\section{Summary}
We have reviewed analytic work on the solutions to the MSW equations
for three neutrino flavours.
Such results are very valuable for a fast scanning over the parameters
of some given model for the mass matrix.

Also, we have more briefly reviewed the LNW mass matrix,
as applied to the neutrino data.
This is a very constrained model that in the quark sector
describes the CKM matrix, and in the neutrino sector gives
the mixing in terms of the mass eigenvalues.

The solar neutrino data has also been studied within the same model,
using numerical integration methods (no ${}_2F_2$'s) \cite{Osland:2000gi}.
An additional fit was then found at $m_1\simeq 2.8\times10^{-6}$~eV,
corresponding to the small-mixing-angle solutions.
However, this point is disfavoured by the atmospheric neutrino data,
and by the electron recoil spectrum.
\bigskip

\leftline{\bf Acknowledgments}
\par\noindent
This work is supported in part by the United
States Department of Energy under Grant No.\ DE-FG02-84ER40158,
and by the Research Council of Norway.

\end{multicols}
\end{document}